%% file: main.tex
\documentclass{article}

    \PassOptionsToPackage{numbers, sort&compress}{natbib}

   \PassOptionsToPackage{numbers,sort&compress}{natbib}
 \usepackage[sglblindworkshop, final]{neurips_2025}
\workshoptitle{MMU-RAG NeurIPS 2025 Competition}

\usepackage[utf8]{inputenc} 
\usepackage[T1]{fontenc}    
\usepackage{hyperref}       
\usepackage{url}            
\usepackage{booktabs}       
\usepackage{amsfonts}       
\usepackage{nicefrac}       
\usepackage{microtype}      
\usepackage{acronym}
\usepackage{graphicx}
\usepackage{fontawesome}

\title{RMIT--ADM+S at the \\MMU-RAG NeurIPS 2025 Competition}

\author{%
  Kun Ran\\
  RMIT University\\ 
  Melbourne, Australia \And
  Marwah Alaofi\\
  RMIT University\\ 
  Melbourne, Australia  \And
  Danula Hettiachchi\\
  RMIT University\\ 
  Melbourne, Australia  \And
  Chenglong Ma\\
  RMIT University\\ 
  Melbourne, Australia  \And
  Khoi Nguyen Dinh Anh\\
  RMIT University\\ 
  Melbourne, Australia  \And
  Khoi Vo Nguyen\\
  RMIT University\\ 
  Melbourne, Australia  \And
  Sachin Pathiyan Cherumanal\\
  RMIT University\\ 
  Melbourne, Australia  \And
  Lida Rashidi\\
  RMIT University\\ 
  Melbourne, Australia  \And
  Falk Scholer\\
  RMIT University\\ 
  Melbourne, Australia  \And
  Damiano Spina\\
  RMIT University\\ 
  Melbourne, Australia \And
  Shuoqi Sun\\
  RMIT University\\ 
  Melbourne, Australia \And
  Oleg Zendel\\
  RMIT University\\ 
  Melbourne, Australia
}

\input{macros}
\input{acronyms}

\begin{document}

\maketitle

\input{sec-abstract}
\input{sec-intro}
\input{sec-qualEval}

\input{sec-system}
\input{sec-conclusions}

\acksection
\vspace*{-0.2cm}
We thank the MMU-RAG organizers for organizing the competition and the
associated awards. This work was supported by the ARC Centre of Excellence for
Automated Decision-Making and Society (ADM+S, CE200100005) and was undertaken
with the assistance of computing resources from RACE (RMIT AWS Cloud
Supercomputing). This system was designed and developed on the unceded lands of
the Woi wurrung and Boon wurrung language groups of the eastern Kulin Nation. We
pay our respects to Elders past and present and extend that respect to all
Aboriginal and Torres Strait Islander peoples and their ongoing connections to
land, sea, sky, and community.

\newpage

\bibliographystyle{plainnat}
\bibliography{references}

\input{appendix}

\end{document}

%% file: macros.tex
\usepackage{xspace}
\usepackage{xcolor}
\usepackage{amsmath}

\makeatletter
\newsavebox\myboxA
\newsavebox\myboxB
\newlength\mylenA

\newcommand*\xoverline[2][0.75]{%
    \sbox{\myboxA}{$\m@th#2$}%
    \setbox\myboxB\null
    \ht\myboxB=\ht\myboxA%
    \dp\myboxB=\dp\myboxA%
    \wd\myboxB=#1\wd\myboxA
    \sbox\myboxB{$\m@th\overline{\copy\myboxB}$}
    \setlength\mylenA{\the\wd\myboxA}
    \addtolength\mylenA{-\the\wd\myboxB}%
    \ifdim\wd\myboxB<\wd\myboxA%
       \rlap{\hskip 0.5\mylenA\usebox\myboxB}{\usebox\myboxA}%
    \else
        \hskip -0.5\mylenA\rlap{\usebox\myboxA}{\hskip 0.5\mylenA\usebox\myboxB}%
    \fi}

\NewDocumentCommand{\sare}{o}{%
	\IfNoValueTF{#1}{\ac{sARE}}{\ac{sARE(#1)}}
}
\NewDocumentCommand{\smare}{o}{%
	\IfNoValueTF{#1}{\ac{sMARE}}{\ac{sMARE(#1)}}
}

\newcommand{\myurl}[1]{{\url{#1}}}

\newcommand{\myparagraph}[1]{\vspace{0.05\baselineskip}\noindent{\textbf{#1.}}}

\newcommand{\mycomment}[1]{}

\definecolor{darkred}{HTML}{B00000}

\newcommand{\prompt}[1]{\begin{quote}\ttfamily #1\end{quote}}

%% file: acronyms.tex
\acrodef{AP}[AP]{Average Precision}
\acrodef{MAP}[MAP]{Mean Average Precision}
\acrodef{nDCG}[nDCG]{normalized Discounted Cumulated Gain}
\acrodef{DCG}[DCG]{Discounted Cumulated Gain}
\acrodef{RR}[RR]{Reciprocal Rank}
\acrodef{MRR}[MRR]{Mean Reciprocal Rank}
\acrodef{ERR}[ERR]{Expected Reciprocal Rank}

\acrodef{QPP}[QPP]{Query Performance Prediction}
\acrodef{QVPP}[QVPP]{Query Variation Performance Prediction}
\acrodef{IR}[IR]{Information Retrieval}
\acrodef{PRP}[PRP]{Probability Ranking Principle}
\acrodef{CIFF}[CIFF]{Common Index File Format}
\acrodef{QL}[QL]{Query Likelihood}
\acrodef{NIR}[NIR]{Neural \ac{IR}}
\acrodef{AQE}[AQE]{Automatic Query Expansion}
\acrodef{BOW}[BoW]{Bag of Words}
\acrodef{MLE}[MLE]{Maximum Likelihood Estimation}
\acrodef{MRF}[MRF]{Markov Random Field}
\acrodef{PRF}[PRF]{Pseudo Relevance Feedback}

\acrodef{QF}[QF]{Query Feedback}

\acrodef{HIT}[HIT]{Human Intelligent Task}
\acrodef{ANOVA}[ANOVA]{ANalysis Of VAriance}
\acrodef{HSD}[HSD]{Honestly Significant Difference}
\acrodef{LtR}[LtR]{Learning to Rank}
\acrodef{NLP}[NLP]{Natural Language Processing}
\acrodef{RM}[RM]{Relevance Model}
\acrodef{LM}[LM]{Language Model}
\acrodef{ML}[ML]{Machine Learning}
\acrodef{SERP}[SERP]{Search Engine Result Page}
\acrodef{RBO}[RBO]{Rank-Biased Overlap}
\acrodef{RBP}[RBP]{Rank-Biased Precision}
\acrodef{ASL}[ASL]{Achieved Significance Level}
\acrodef{RSV}[RSV]{Retrieval Status Value}
\acrodef{DFR}[DFR]{Divergence From Randomness}
\acrodef{PMI}[PMI]{Pointwise Mutual Information}

\acrodef{CLT}[CLT]{Central Limit Theorem (CLT)}
\acrodef{iid}[i.i.d]{independent and identically distributed}
\acrodef{CI}[CI]{Confidence Interval}
\acrodef{MAE}[MAE]{Mean Absolute Error}
\acrodef{MSE}[MSE]{Mean Squared Error}
\acrodef{RV}[RV]{Random Variable}
\acrodef{GLM}[GLM]{General Linear Model}
\acrodef{LLM}[LLM]{Large Language Model}
\acrodef{GLMM}[GLMM]{General Linear Mixed Model}
\acrodef{GPT}[GPT]{Generative Pre-trained Transformer}
\acrodef{SOA}[SOA]{Strength of Association}
\acrodef{IQR}[IQR]{Interquartile Range}
\acrodef{BCA}[BCa]{Bias-Corrected and accelerated}
\acrodef{FWER}[FWER]{Family-Wise Error Rate}
\acrodef{KS}[KS]{Kolmogorov-Smirnov}
\acrodef{SD}[SD]{Standard Deviation}
\acrodef{SS}[SS]{Sum of Squares}

\acrodef{sMARE(AP)}[sMARE-AP]{AP induced scaled Mean Absolute Rank Error}
\acrodef{sMARE}[sMARE]{scaled Mean Absolute Rank Error}
\acrodef{sARE}[sARE]{scaled Absolute Rank Error}
\acrodef{sARE(AP)}[sARE-AP]{AP induced scaled Absolute Rank Error}
\acrodef{CDF}[CDF]{Cumulative Density Function}
\acrodef{eCDF}[eCDF]{Empirical Cumulative Density Function}
\acrodef{KDE}[KDE]{Kernel Density Estimation}
\acrodef{SOTA}[SOTA]{State Of The Art}
\acrodef{KLD}[KLD]{Kullback-Leibler Divergence}
\acrodef{KL}[KL]{Kullback-Leibler}
\acrodef{PDF}[PDF]{Probability Density Function }
\acrodef{RMSE}[RMSE]{Root Mean Square Error}

\acrodef{TREC}[TREC]{Text REtrieval Conference}
\acrodef{NIST}[NIST]{National Institute of Standards and Technology}
\acrodef{CLEF}[CLEF]{Conference and Labs of the Evaluation Forum}
\acrodef{NTCIR}[NTCIR]{NII Testbeds and Community for Information access Research}
\acrodef{FIRE}[FIRE]{Forum for Information Retrieval Evaluation}
\acrodef{INEX}[INEX]{Initiative for the Evaluation of XML Retrieval}
\acrodef{WWW}[WWW]{World Wide Web}
\acrodef{RAG}[RAG]{Retrieval-Augmented Generation}

\acrodef{LR}[$LR$]{Logistic Regression}

%% file: sec-abstract.tex
\begin{abstract}
    This paper presents the award-winning RMIT-ADM+S system for the Text-to-Text
    track of the NeurIPS~2025 MMU-RAG Competition. We introduce Routing-to-RAG
    (R2RAG), a research-focused retrieval-augmented generation (RAG)
    architecture composed of lightweight components that dynamically adapt the
    retrieval strategy based on inferred query complexity and evidence
    sufficiency. The system uses smaller LLMs, enabling operation on a single
    consumer-grade GPU while supporting complex research tasks. It builds on the
    G-RAG system, winner of the ACM~SIGIR~2025 LiveRAG Challenge, and extends it
    with modules informed by qualitative review of outputs. R2RAG won the Best
    Dynamic Evaluation award in the Open Source category, demonstrating high
    effectiveness with careful design and efficient use of resources.
\end{abstract}

%% file: sec-intro.tex
\section{Introduction}
\label{sec:intro}

\acf{RAG}~\cite{lewis2020retrieval} has become a standard approach for improving
the grounding and reliability of \acp{LLM}. Evaluation campaigns such as
LiveRAG~\cite{carmel2025sigir2025liverag} provide standardized settings for
deploying and assessing \ac{RAG} systems. MMU-RAG, a NeurIPS~2025 competition
with a focus on user-centric evaluation, complements these efforts. Final
rankings were determined across four categories: \textit{Static} and
\textit{Dynamic} evaluation, and \textit{Open Source} and \textit{Closed Source}
system types. The evaluation framework combined a robustness-aware aggregation
of normalized automatic metrics and human judgments based on ordinal
Likert-scale ratings, spanning both static and real-time dynamic (RAG-Arena)
evaluation modes.\footnote{See MMU-RAG:
\url{https://agi-lti.github.io/MMU-RAGent/}}  
Our system, R2RAG, was a top-performing submission in the Text-to-Text track,
excelling in the open-source dynamic evaluation mode.

Recent research on \ac{RAG} has increasingly emphasized the importance of
user-centric perspective by making \ac{RAG} systems more adaptive, enabling them
to plan searches, inspect retrieved evidence, and adjust their actions
dynamically throughout a task. Motivated by the real-time evaluation setting of
MMU-RAG, our system is designed to support dynamic decision-making through
specialized agents that manage planning, search, and reasoning. The goal is to
construct a clear and efficient pipeline that can run on low-cost GPUs and
consumer-grade hardware while addressing complex information needs. To
facilitate reproducibility, we release our code repository.\footnote{See
\url{https://github.com/rmit-ir/NeurIPS-MMU-RAG}}

This report presents our system for the MMU-RAG Text-to-Text track. We describe
the architecture, justify the design choices behind each component, and explain
how insights from a qualitative evaluation study informed system refinement. 

%% file: sec-qualEval.tex
\section{Qualitative Evaluation}
\label{sec:qual-eval}

To study the strengths and limitations of different retrieval and generation
strategies, we conducted a qualitative focus-group evaluation. The session
examined differences in retrieval quality, synthesis coherence, response
relevance, verbosity, and overall perceived usefulness in (semi-)realistic
deep-research scenarios. Consistent with Section~\ref{sec:intro}, the goal was
to assess user-perceived trade-offs that may not appear in aggregate metrics.

The initial system was adapted from our champion submission to the SIGIR 2025
LiveRAG~\cite{carmel2025sigir2025liverag} challenge, which relied on synthetic
LLM-generated queries and LLM-as-a-Judge evaluation~\cite{ran2025rmit}. For
MMU-RAG, we focused on performance under the RAG Arena framework. While
synthetic evaluation enables rapid iteration, prior work highlights limitations,
including distributional mismatch and evaluation
biases~\cite{zendel2025comparative,dietz2025principles,rahmani2025biassyntheticdata,balog2025rankers}.
To capture realistic user interactions and failure modes, we emphasized
qualitative assessment.

We first involved the co-authors as representative users and later extended the
session to a broader group from RMIT University's Centre for Human-AI
Information Environments (CHAI).\footnote{See
\url{https://www.rmit.edu.au/chai}} The study included ($N=20$) participants
across academic levels, from master's students to professors, with expertise in
information retrieval, machine learning, human-computer interaction, natural
language processing, and data science.\footnote{No personal or identifiable data
were collected or stored; feedback was used solely to improve the system and was
not shared externally.}

Participants interacted with the system by submitting queries and providing
binary feedback (\faThumbsUp/\faThumbsDown), open-ended comments, and verbal
reflections. The two-hour session allowed free exploration, during which
participants submitted an average of 17 queries.

We analyzed feedback for each query using two complementary approaches: (i)
Preference Ratio, computed as the ratio of \faThumbsUp~ to \faThumbsDown~
responses, and (ii) manual inspection of open-ended comments to identify
qualitative themes and contextual nuances. The diversity of feedback proved
informative in two respects. First, it enabled us to identify concrete system
limitations, including failure cases in retrieval grounding and synthesis.
Second, it surfaced dimensions that mattered differently to participants, such
as perceived system biases, preferences for response structure and level of
detail, and sensitivity to tone and framing. These aspects were not directly
captured by quantitative metrics but influenced overall user judgments. Based on
these findings, we refined several prompts and tuned selected system components
to address recurring issues and improve alignment with user expectations.

Although the Preference Ratio suggested similar overall performance between the
two variants, qualitative analysis revealed complementary strengths: Vanilla
Agent handled complex queries more effectively but was verbose for simple ones,
whereas Vanilla RAG was concise for simple queries but weaker on multi-faceted
tasks.

These observations align with the MMU-RAG organizers' key insight:
\smallskip
\begin{quote}
    ``\textbf{Live evaluation matters:} Arena preferences revealed qualitative
    distinctions not captured by static metrics.''
\end{quote}
\smallskip
Our findings likewise indicate that preference-based and qualitative feedback
can reveal system trade-offs that are not apparent from aggregate metrics alone.

%% file: sec-system.tex
\section{System Architecture}
\label{sec:system}

\begin{figure}[bt]
    \centering
    \includegraphics[width=\linewidth]{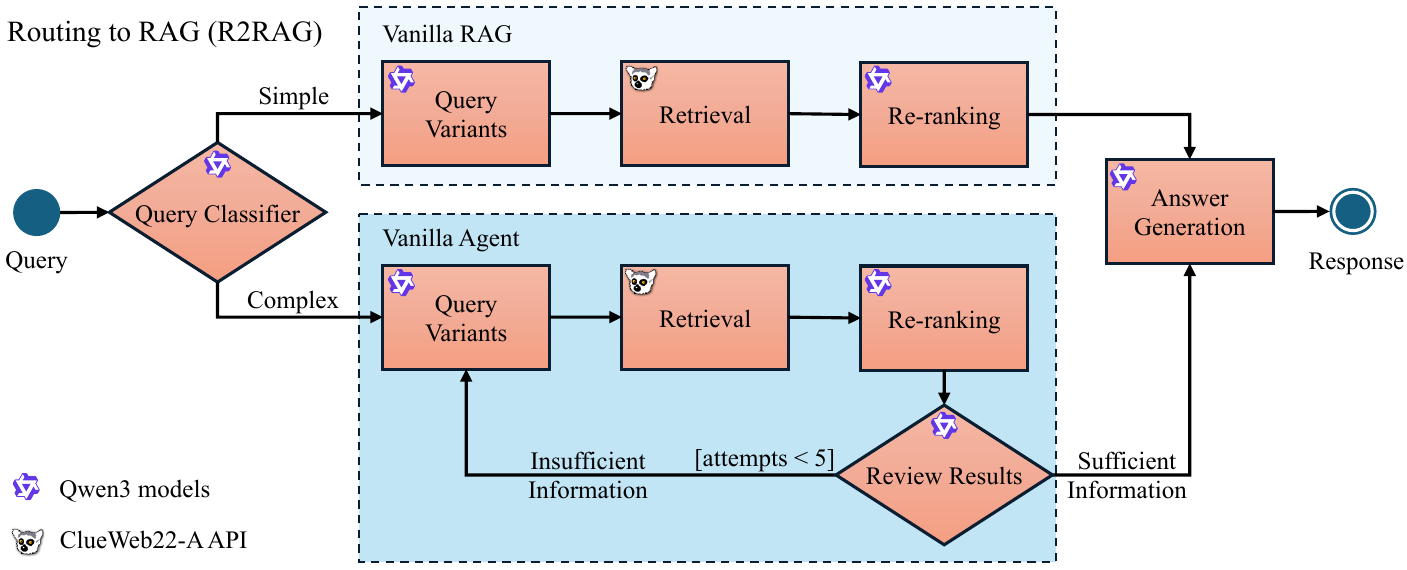}
    \caption{Overview of the system. The workflow includes a query classifier
        that routes each query to either a Vanilla RAG path for simple queries
        or a Vanilla Agent for complex queries that iteratively issues revised
        search query variants until sufficient evidence is gathered, followed by
        an answer generation module.}
    \label{fig:agent_workflow}
\end{figure}

A central design principle of R2RAG is the explicit distinction between simple
and complex queries, as illustrated in Figure~\ref{fig:agent_workflow}. The
system first applies a query classifier that assigns each incoming query to one
of these two categories. Based on this decision, the query is routed to either a
\textit{Vanilla RAG} pipeline for simple queries or a \textit{Vanilla Agent}
pipeline for complex queries. Simple queries are processed using a single
retrieval step followed by answer generation. In contrast, complex queries are
handled through an iterative process in which the agent reformulates search
queries and performs multiple retrieval rounds to gather complementary and
sufficient evidence. After the relevant documents are collected, an
\ac{LLM}-based answer generator produces the final response.

Our system relies on \ac{LLM}-based inference for query classification, query
reformulation, document assessment, and answer generation. Model selection was
guided by model size, available GPU memory, context window capacity, and
reasoning performance under resource constraints.

We used the \texttt{Qwen3-4B}~\cite{yang2025qwen3} \ac{LLM} for most components
and \texttt{Qwen3-Reranker-0.6B}~\cite{qwen3embedding} for document reranking.
The 4B model was used without quantization, as prior work shows that
quantization can degrade long-context reasoning performance in
\acp{LLM}~\cite{liu-etal-2024-emergent,mekala-etal-2025-quantization}. In
preliminary experiments, 4-bit quantized larger models (e.g., \texttt{Qwen3-8B})
yielded weaker performance than the unquantized 4B variant. The compact size of
\texttt{Qwen3-4B} allows it to fit within the available GPU memory alongside the
reranker model, while maintaining strong performance across classification and
generation tasks.

Both models were served using vLLM~\cite{kwon2023efficient} to support efficient
inference. To balance efficiency and performance, \texttt{Qwen3-4B} was
configured with a reduced context window of 25{,}000 tokens (from the native
32{,}768 tokens) and run with the developer-recommended decoding parameters,
\emph{temperature} = 0.6 and \emph{top\_p} = 0.95.

\subsection{Query Classifier}

Queries vary in complexity and therefore require different retrieval strategies.
We classify queries as \emph{simple} or \emph{complex}, where complex queries
are defined as non-factoid questions that require long-form reasoning or
multi-step synthesis~\cite{rosset2025researchyquestions}. We explored two
classifier approaches:

\myparagraph{Logistic Regression (LR) Classifier} We trained and evaluated the LR
classifier on a dataset of 175{,}850 queries combining TREC Deep
Learning~\cite{craswell2021trecdl}, Deep-Research
Questions~\cite{rosset2025researchyquestions}, TREC RAG
2025~\cite{thakur2025assessingtrecrag2025,pradeep2025ragnarok}, and Natural
Questions~\cite{kwiatkowski2019naturalquestions}. Queries requiring
decomposition were labeled \texttt{True} (complex), and others \texttt{False}
(simple). Each query was represented by a 147-dimensional feature vector,
consisting of a 128-dimensional semantic embedding and 19 linguistic
features,\footnote{Using the \texttt{google/bert\_uncased\_L-2\_H-128\_A-2}
    model from Hugging Face.} including part-of-speech and punctuation
counts.\footnote{Using the English language model \texttt{en\_core\_web\_sm}
    from spaCy.}

\myparagraph{LLM-based Classifier} We used a structured prompt
(Appendix~\ref{sec:appendix_query_classifier}) to instruct the \ac{LLM} to
classify queries as complex (requiring more than a single Google search) or
simple. This operational definition differs from standard \ac{IR} definitions
but is appropriate for routing queries to the corresponding retrieval path.

The LLM-based classifier offers several advantages over LR. It can reason about
query semantics, better handle out-of-distribution queries, and allows flexible
adjustment of classification criteria through prompt modifications without
retraining. The main trade-off is higher computational cost, as \ac{LLM}
inference requires more time and GPU resources. Given the 10-minute per-query
limit, we prioritized these advantages over the efficiency of the LR model and
adopted the LLM-based classifier, as the additional overhead remains acceptable
within this constraint.

\subsection{Vanilla RAG}

For queries classified as simple, we use a \ac{RAG} pipeline called
\textit{Vanilla RAG}, which processes the query in a single pass without
iterative refinement. As shown in the top path of
Figure~\ref{fig:agent_workflow}, the pipeline includes four stages:

\myparagraph{Generating Query Variants} To increase retrieval coverage, we generate
three query variants using Qwen3-4B with thinking mode enabled. Following
\citet{ran2025twoheads}, these variants rephrase the original query with
different keywords and formulations to capture a broader range of relevant
documents.

\myparagraph{Retrieval} Each variant is used to search the ClueWeb22-A index in
parallel, returning up to ten documents per query.\footnote{Search was performed
    through the API provided by the organizers:
    \url{https://www.deepresearchgym.ai/}} After deduplication, this yields fewer
than 30 documents for the next stage.

\myparagraph{Reranking} The aggregated search results are reranked using the
Qwen3-reranker-0.6b model with customized instructions (instruction provided in
Appendix \ref{sec:appendix_reranker}).\footnote{Model card:
    \url{https://huggingface.co/Qwen/Qwen3-Reranker-0.6B}} The model serves as a
Pointwise reranker~\cite{pointwise2024reranking}, predicting ``yes'' or ``no''
for each query-document pair and using the probability of ``yes'' as the
relevance score. We chose this model because its size fits the hardware budget
while allowing the rest of the system to run efficiently. After reranking, we
select the top-ranked documents and truncate them to a 5k-word limit to balance
coverage and computational cost.

\myparagraph{Answer Generation} The reranked documents and the original query are
passed to Qwen3-4B for answer generation. The model is instructed to synthesize
information from the retrieved documents, produce a coherent response suited to
the query, and include inline citations in the format [ID] to reference specific
source documents. For controversial topics, the model is encouraged to provide
balanced perspectives. The full prompt is provided in
Appendix~\ref{sec:appendix_answer_generator}. Note that retrieval and reasoning
are performed over entire documents rather than chunks.

\subsection{Vanilla Agent}
\label{sec:vanilla_agent}

For complex queries, we developed \textit{Vanilla Agent}, an iterative \ac{RAG}
system that performs multiple rounds of retrieval to gather complementary
information. As shown in the lower path of Figure~\ref{fig:agent_workflow}, it
shares the Query Variants, Retrieval, and Reranking components with Vanilla RAG
but operates within an iterative search loop with enhanced parameters.

\myparagraph{Iterative Search Loop} Each iteration runs the Query
Variants-Retrieval-Reranking pipeline with more aggressive settings, generating
up to five query variants (instead of three) and allowing a larger document
limit of up to 25K tokens (compared to 5K words).\footnote{Tokens are used here
to ensure the model stays within its context limits, while Vanilla RAG uses
words for faster estimation.} The system maintains state across iterations by
accumulating: (1) previously used queries to avoid repetition, and (2) summaries
of useful documents to guide subsequent searches.

The stopping condition $S$ depends on the accumulated token count $(T)$,
estimated information coverage $(Cov)$, and a fixed iteration cap $(i)$:
$$\text{S} \iff (T > 20{,}000) \ \lor \ (Cov = 1) \ \lor \ (i \ge 5).$$
The loop terminates when any of these conditions is met. The token threshold
protects against exceeding the model context window during final answer
generation while indicating sufficient information density. The coverage signal
is determined by the \ac{LLM} based on the accumulated agent state. The
iteration cap prevents unbounded search loops.

\myparagraph{Document Review and Information Coverage} The $Cov$ factor is
updated after each reranking step by the document review component, using
\texttt{Qwen3-4B} (prompt in Appendix~\ref{sec:appendix_review_results}). The
model evaluates: (1) whether the query requires coverage of multiple
perspectives, (2) whether all sub-parts of the query are addressed, (3) balance
for contested topics, (4) identification of newly useful documents, and (5)
remaining knowledge gaps. If the accumulated information is insufficient, the
model generates a new query targeting the identified gaps while avoiding
previous unsuccessful queries. If no documents are retrieved in an iteration,
the system reformulates the query and proceeds to the next iteration.

\subsection{Deployment and Evaluation Infrastructure}
\vspace*{-0.2cm}
During development, we implemented our different \ac{RAG} system variants as
services that can be deployed through multiple API servers, including the
MMU-RAG API server required by the competition guidelines and an
OpenAI-compatible API server that exposes multiple \ac{RAG} variants on a single
GPU server for our evaluation infrastructure. Each system configuration is
assigned a unique model identifier in the \texttt{model} field. This design
eliminates the need to deploy separate servers for each variant, thereby
simplifying infrastructure management. For internal qualitative testing, we
integrated the API with an in-house search engine that allows users to submit
queries to different \ac{RAG} systems and provide feedback for the responses.

%% file: sec-conclusions.tex
\section{Conclusions}
\label{sec:conclusions}

This paper presented R2RAG, our dynamic \ac{RAG} system for the Text-to-Text
track of the NeurIPS 2025 MMU-RAG Competition. The system adapts its retrieval
strategy based on query complexity, routing queries to either a single-pass
Vanilla RAG pipeline or an iterative Vanilla Agent that accumulates evidence
across multiple retrieval rounds. For complex queries, an \ac{LLM}-based review
module estimates information coverage and guides query reformulation, with
termination governed by coverage signals, token budget, and iteration limits.

Our results show that small reasoning-oriented models can effectively support
dynamic \ac{RAG} pipelines under realistic resource constraints. Retrieval
quality and controlled evidence accumulation were central to answer quality,
while structured \ac{LLM}-based decision components improved robustness. The
entire system operates on a single consumer-grade GPU through compact model
selection and careful resource tuning.

Beyond system design, our experience highlights the importance of well-aligned,
multi-level evaluation. Static automatic metrics provided useful baselines but
did not fully capture qualitative differences in reasoning depth, presentation,
and user preference. Qualitative and dynamic evaluation settings were essential
for identifying failure modes, refining prompts, and validating system behavior
under realistic conditions. Strong performance in the competition was therefore
not only a result of architectural choices, but also of rigorous, experimentally
grounded evaluation that aligned testing conditions with real user needs.

%% file: appendix.tex
\appendix
\label{sec:appendix}

\section{Prompts}
\label{sec:appendix_prompts}

\subsection{Query Classifier} \label{sec:appendix_query_classifier}

\paragraph{System Prompt:}

\prompt{Judge if the user question is a complex question. Note that the answer
can only be "yes" or "no".

Given the question below, if you are doing the research, do you think the
question is very easy and you can find the answer easily with a single search on
Google?

If so, it's not a complex question, respond with "no", otherwise, it's a complex
question, respond with "yes".

Generally, for straightforward, single question, it's a simple question. If the
question is ambiguous, multifaceted, contains multiple parts, has 2 or more
sub-questions or requires multiple steps to answer, it's a complex question.

Give the final answer based on your last reasoning, yes indicates it's a complex
question or no indicating it's not a complex question.}

\paragraph{User Prompt:}

\prompt{\{question\}}

\subsection{Query Variants Generator} \label{sec:appendix_query_variants_generator}

\paragraph{System Prompt:}

\prompt{You will receive a question from a user and you need interpret what the
question is actually asking about and come up with 2 to 5 Google search queries
to answer that question.

Try express the same question in different ways, use different techniques, query
expansion, query relaxation, query segmentation, use different synonyms, use
reasonable guess and different keywords to reach different aspects.

Try to provide a balanced view for controversial topics.

To comply with the format, put your query variants enclosed in queries xml
markup:

<queries>\\
query variant 1\\
query variant 2\\
...\\
</queries>

Put each query in a line, do not add any prefix on each query, only provide the
query themselves.}

\paragraph{User Prompt:}

\prompt{User question: \{query\}}

\subsection{Re-ranker} \label{sec:appendix_reranker}

\paragraph{Prompt:}

\prompt{<|im\_start|>system

Judge whether the Document meets the requirements based on the Query and the Instruct provided. Note that the answer can only be "yes" or "no".<|im\_end|>

<|im\_start|>user

<Instruct>: Given the web search query, is the retrieved document\\
(1) from a high quality and relevant website based on the URL,\\
(2) published recently, and\\
(3) contains key information that helps answering the query?

<Query>: \{query\}

<Document>: Web URL: \{url\}

Content: \{document\_text\}

<|im\_end|>

<|im\_start|>assistant

<think>\\
</think>}

\subsection{Review Results} \label{sec:appendix_review_results}

\paragraph{System Prompt:}

\prompt{You are an expert in answering user question "\{question\}". We are doing research on user's question and currently working on aspect "\{next\_query\}"

Go through each document in the search results, judge if they are sufficient for answering the user question. Please consider:

1. Does the user want a simple answer or a comprehensive explanation? For comprehensive explanation, we may need searching with different aspects to cover a wider range of perspectives.\\
2. Does the search results fully addresses the user's query and any sub-components?\\
3. For controversial, convergent, divergent, evaluative, complex systemic, ethical-moral, problem-solving, recommendation questions that will benefit from multiple aspects, try to search from different aspects to form a balanced, comprehensive view.\\
4. Don't mention irrelevant, off-topic documents.\\
5. When you answer 'yes' in <is-sufficient>, we will proceed to generate the final answer based on these results. If you answer 'no', we will continue the next turn of using your new query to search, and let you review again.\\
6. If information is missing or uncertain, always return 'no' in <is-sufficient> xml tags for clarification, and generate a new query enclosed in xml markup <new-query>your query</new-query> indicating the clarification needed. If the search results are too off, try clarifying sub-components of the question first, or make reasonable guess. If you think the search results are sufficient, return 'yes' in <is-sufficient> xml tags.\\
7. If current search results have very limited information, try use different techniques, query expansion, query relaxation, query segmentation, use different synonyms, use reasonable guess and different keywords to get relevant results, and put the new query in <new-query>your query</new-query> xml tags. If there are previous search queries, do not repeat them in the new query, we know they don't work.\\
8. Identify unique, new documents that are important for answering the question but not included in previous documents, and list their IDs (\# in ID=[\#]) in a comma-separated format within <useful-docs> xml tags. If multiple documents are similar, choose the one with better quality. Do not provide duplicated documents that have been included in previous turns. If no new documents are useful, leave <useful-docs></useful-docs> empty.\\
9. If useful-docs is not empty, provide a brief summary of what these documents discuss within <useful-docs-summary> xml tags, in 1-2 sentences. Start your summary with "These documents discuss...". Do not mention specific document IDs in the summary.\\
10. Your purpose is to judge documents relevance against the question, not to provide the final answer yet, do not answer the question.

Response format:

- <is-sufficient>yes or no</is-sufficient> (For all of the documents we have collected, including previous documents, do we have enough information to answer the user question? If yes, then provide <useful-docs> and <useful-docs-summary> tags; if 'no', then provide <new-query> tag)\\
- <new-query>your new query</new-query> (only if is-sufficient is 'no')\\
- <useful-docs>1,2,3</useful-docs> (list of document IDs that are useful for answering the question, separated by commas)\\
- <useful-docs-summary></useful-docs-summary> (short summary of what these useful documents are talking about, just the summary, only if useful-docs is not empty)\\

\{prev\_questions\}\\
\{prev\_docs\_summaries\}\\
Here is the current question: "\{query\}"\\
Here is the search results for current question:

<search-results>\\
Webpage ID=[1] URL=[\{url1\}] Date=[\{date1\}]:

\{document\_text\}

Webpage ID=[2] URL=[\{url2\}] Date=[\{date2\}]:

\{document\_text\}

</search-results>}

\subsection{Answer Generator} \label{sec:appendix_answer_generator}

\paragraph{System Prompt:}

\prompt{You are a knowledgeable AI search assistant built by the RMIT IR team.

Your search engine has returned a list of relevant webpages based on the user's question, listed below in <search-results> tags. These webpages are your knowledge.

The next user message is the full user question, and you need to explain and answer the question based on the search results. Do not make up answers that are not supported by the search results. If the search results do not have the necessary information for you to answer the search question, say you don't have enough information for the question.

Try to provide a balanced view for controversial topics.

Tailor the complexity of your response to the user question, use simpler bullet points for simple questions, and sections for more detailed explanations for complex topics or rich content.

Do not answer to greetings or chat with the user, always reply in English.

You should refer to the search results in your final response as much as possible, append [ID] after each sentence to point to the specific search result. e.g., "This sentence is referring to information in search result 1 [1].".

Current time at UTC+00:00 timezone: \{datetime.now(timezone.utc)\}\\
Search results knowledge cutoff: December 2024

<search-results>\\
Webpage ID=[1] URL=[\{url1\}] Date=[\{date1\}]:

\{document\_text\}

Webpage ID=[2] URL=[\{url2\}] Date=[\{date2\}]:

\{document\_text\}

</search-results>}

\paragraph{User Prompt:}

\prompt{\{question\}}